# Modified interlayer stacking and insulator to correlated-metal transition driven by uniaxial strain in 1$T$-TaS$_2$

Christopher W. Nicholson[1,2,*], Francesco Petocchi[1], Björn Salzmann[1], Catherine Witteveen[3,4], Maxime Rumo[1], Geoffroy Kremer[1], Fabian O. von Rohr[3,4], Philipp Werner[1], Claude Monney[1]

[1]University of Fribourg and Fribourg Centre for Nanomaterials, Chemin du Musée 3, CH-1700, Fribourg, Switzerland

[2]Fritz Haber Institute of the Max Planck Society, Faradayweg 4-6, D-14195, Berlin, Germany

[3]Department of Physics, University of Zurich, Winterthurerstrasse 190, CH-8057 Zurich, Switzerland

[4]Department of Quantum Matter Physics, University of Geneva, 24 Quai Ernest-Ansermet, CH-1211 Geneva, Switzerland

**Interlayer coupling is strongly implicated in the complex electronic properties of 1$T$-TaS$_2$, but the interplay between this and electronic correlations remains unresolved. Here, we employ angle-resolved photoemission spectroscopy (ARPES) to reveal the effect of uniaxial strain engineering on the electronic structure and interlayer coupling in 1$T$-TaS$_2$. The normally insulating ground state is transformed into a correlated-metal phase under strain, as evidenced by the emergence of a narrow band at the Fermi level. Temperature dependent ARPES measurements reveal that the metallic behaviour only develops below the commensurate charge density wave (CCDW) transition, where interlayer dimerization produces a band-insulator in unstrained samples. Electronic structure calculations demonstrate that the correlated metallic behaviour is stabilized by a previously predicted but unobserved bulk stacking structure with a modified interlayer coupling of the Ta $d_{z^2}$ electrons. Our combined approach lays bare the role of correlations and interlayer coupling in 1$T$-TaS$_2$, providing critical input for understanding superconductivity under pressure and the metastable hidden phase induced using non-equilibrium protocols in this platform material for correlated physics.**

The overlap between electronic wavefunctions on neighbouring atomic sites determines the kinetic energy, or bandwidth, of electronic states in solids. Reducing this overlap to produce narrowly dispersing electronic bands promotes correlated behaviour in the presence of electronic interactions[1,2]. This can have particularly striking effects in low-dimensional materials, as exemplified by the milestone discoveries of the fractional quantum hall effect[3] and high-$T_c$ superconductivity in the cuprates[4]. In layered (quasi-2D) materials, the out-of-plane coupling is typically weaker than that in the plane. Nevertheless, coupling along the out-of-plane direction can play a crucial role in the overall energetics of a material. This manifests itself in twisted bilayer graphene, where modified interlayer coupling leads to emergent strongly correlated phases[5,6] and flat electronic bands[7]. Indeed, the higher sensitivity of interlayer coupling to external perturbations provides a control parameter to modify correlated electron behaviour. However, twist angle engineering is incompatible with bulk quasi-2D materials, where interlayer coupling is strongly implicated in the rich physics observed. Well-known examples include pressure-induced and enhanced superconductivity in numerous layered compounds and more recently the diverse and intriguing physics of the newly discovered bulk Kagome series $A$V$_3$Sb$_5$[8], hosting diverse phenomena which are tuneable with pressure[9] and with interlayer ordering[10]. Such properties do not necessarily translate to the monolayer limit in the absence of interlayer coupling. Similarly limiting, high-pressure cells exclude access with direct probes of the momentum-resolved electronic structure such as ARPES.

An approach that combines the two is uniaxial strain engineering, which can modify the crystal structure and hence the electronic kinetic energy in the Hamiltonian while not constraining access to the sample surface, a necessary condition for ARPES. This approach has proved highly successful at modifying correlated behaviour in bulk correlated oxides[11–15]. Strain has furthermore been shown to produce equivalent effects compared with moiré engineering[16,17] and therefore offers a route to control interlayer coupling and emergent properties in a broad range of materials. However, little work on strain engineering in quasi-2D correlated materials has been performed, and the combination with ARPES remains challenging and limited[18–21].

Layered 1$T$-TaS$_2$ hosts a particularly rich and controversial phase diagram, and has therefore become an important platform material for studying electronic correlations in low dimensions. In addition to a series of charge density wave (CDW) phases upon cooling – with incommensurate (550 K), nearly-commensurate (350 K) and commensurate (180 K) structures[22] – TaS$_2$ also displays superconductivity under pressure[23], a metastable "hidden" phase following pulsed excitation[24,25], is a candidate quantum spin liquid[26], and hosts a controversial Mott phase in the commensurate CDW (CCDW) phase that has been debated for almost 50 years[27–29]. Recently this debate has intensified[30–34], motivated in part by the fact that doped Mott insulators represent the prototypical framework for understanding high-$T$c superconductivity in the cuprates[35].

The out-of-plane stacking structure of TaS$_2$ along the $c$-axis is irregular in the proposed Mott phase[36,37], with bilayer sheets separated by a randomly assigned off-centre stacking vector[37–40] (see Fig. 2b). An important theoretical step was made when Ritschel[31] and Lee[32] showed that a gapped electronic structure could be obtained via bilayer coupling, as originally suggested by Naito[40]. In a bilayer structure (Fig. 1) a single electron from each layer can hybridize with the corresponding electron in a neighbouring layer and open a bonding-antibonding gap, thus removing the need for Mott correlations. This arises because the "star-of-David" structures generated by the ionic displacements in the CDW phases contain 13 in-plane Ta atoms, with a total of 12 paired electrons and a single unpaired electron on the central Ta atom in a $d_{z^2}$ orbital[41,42] (see Fig. 2a, b). However, this implies that a split bilayer created by e.g. surface termination (Fig. 1, centre) would result in fundamentally different electronic properties[31,33]. Recent STM results confirmed that terminations corresponding to complete and split bilayers are indeed observed[33,34,43]. Moreover both surfaces were found to be gapped; an unexpected result within the single-particle picture. This suggests that at least one termination contains Mott physics, as highlighted by recent calculations[44]. However, in the bulk material any possible Mott correlations are ineffective due to the presence of the hybridisation gap.

This begs the question: what happens to the bulk electronic properties if the interlayer stacking is modified away from the ground state and the bilayer structure is disturbed? This is critical information considering that superconductivity appears when the system is pushed away from standard equilibrium conditions by pressure. Similarly the metallic hidden phase driven by non-equilibrium protocols, for which TaS$_2$ is the prototypical material, may involve stacking rearrangements[25]. Nonetheless, the mechanisms through which these phases occur remain unclear, and direct momentum-resolved electronic structure information is difficult to obtain due to the nature of the techniques involved in producing them. Overall, a definitive statement of the role played by electronic correlations and interlayer stacking in 1$T$-TaS$_2$, and their relation to the various emergent phases in this material, remains elusive.

Here, we follow the modifications of the momentum-resolved electronic structure of 1$T$-TaS$_2$ under uniaxial strain. We demonstrate the appearance of a low-temperature correlated metallic phase that corresponds to a stacking structure not present under standard equilibrium conditions. These results highlight the central role played by interlayer coupling in determining the ground state properties of correlated 1T-TaS$_2$, opening a new route to manipulate electronic properties in this and other layered compounds.

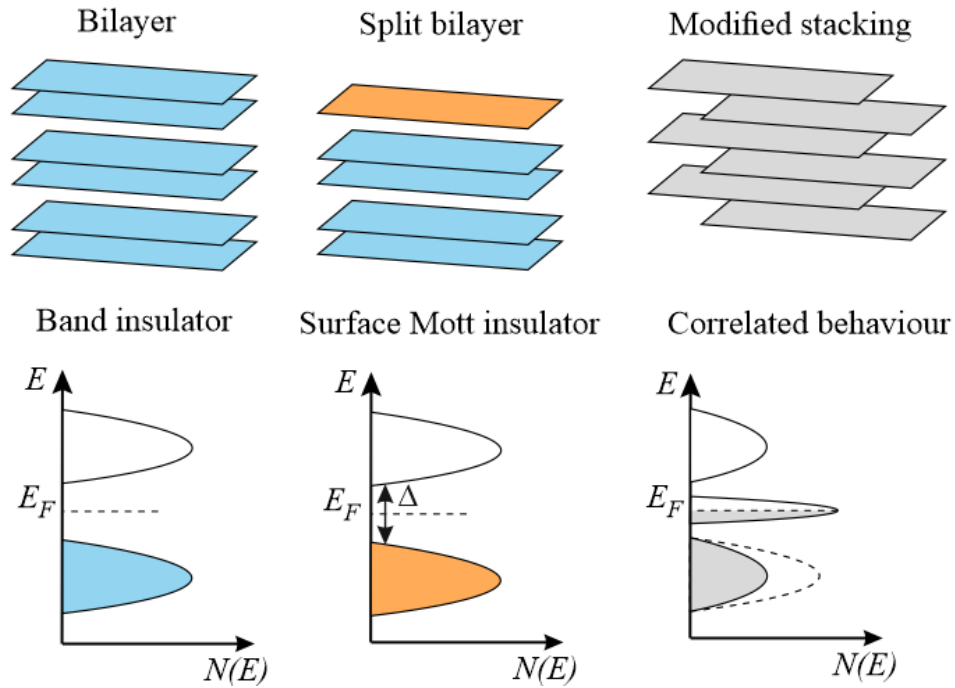

**Fig. 1 Schematic relation between interlayer stacking and electronic structure for 1T-TaS$_2$.** The ground state of TaS$_2$ consists of bilayers along the out of plane direction (see also Fig. 2b) there interlayer hybridisation leads to band insulating behaviour (left panel). If the bilayer is split e.g. at the surface (central panel) the single unpaired electron should be metallic, but due to Mott correlations is also insulating. If the interlayer stacking were to be modified away from the bilayer structure, this may significantly influence the electronic properties. Uniaxial strain can alter atomic site overlap and thus change the electronic landscape such that the stacking is modified.

**Results**

The electronic structure of an unstrained TaS$_2$ crystal in the CCDW phase is presented in Fig. 2c along the $\bar{\Gamma} - \bar{M}$ direction of the surface Brillouin zone (Fig. 2d). The Ta 5$d$ manifold extends from $E_F$ to -1.5 eV, with S 2$p$ states at lower energies [41,42]. These data obtained with 21.2 eV excitation energy are in excellent agreement with previous reports [41,45,46], as are the corresponding transport and structural properties, which are presented in Extended Data Fig. 1.

Utilising a low-energy laser source (see Methods) to focus on the low energy region near the $\bar{\Gamma}$-point we observe a clear splitting of the Ta 5$d$ states due to the CCDW phase (Fig. 2e), again in line with previous observations [47]. However, by introducing uniaxial strain to the same crystal (see Methods), unexpected additional spectral weight appears at $E_F$ in a narrowly dispersing band (Fig. 2f). This is further emphasized by the direct comparison of integrated line cuts presented in Fig. 2g for unstrained and strained samples. A clear quasi-particle peak is evident at $E_F$ only in the strained case, revealing a strain induced insulator to metal transition. As highlighted in Extended Data Fig. 2 the bandwidth of this state is only 70 meV, comparable to the flat bands observed recently in twisted bilayer graphene [7,48].

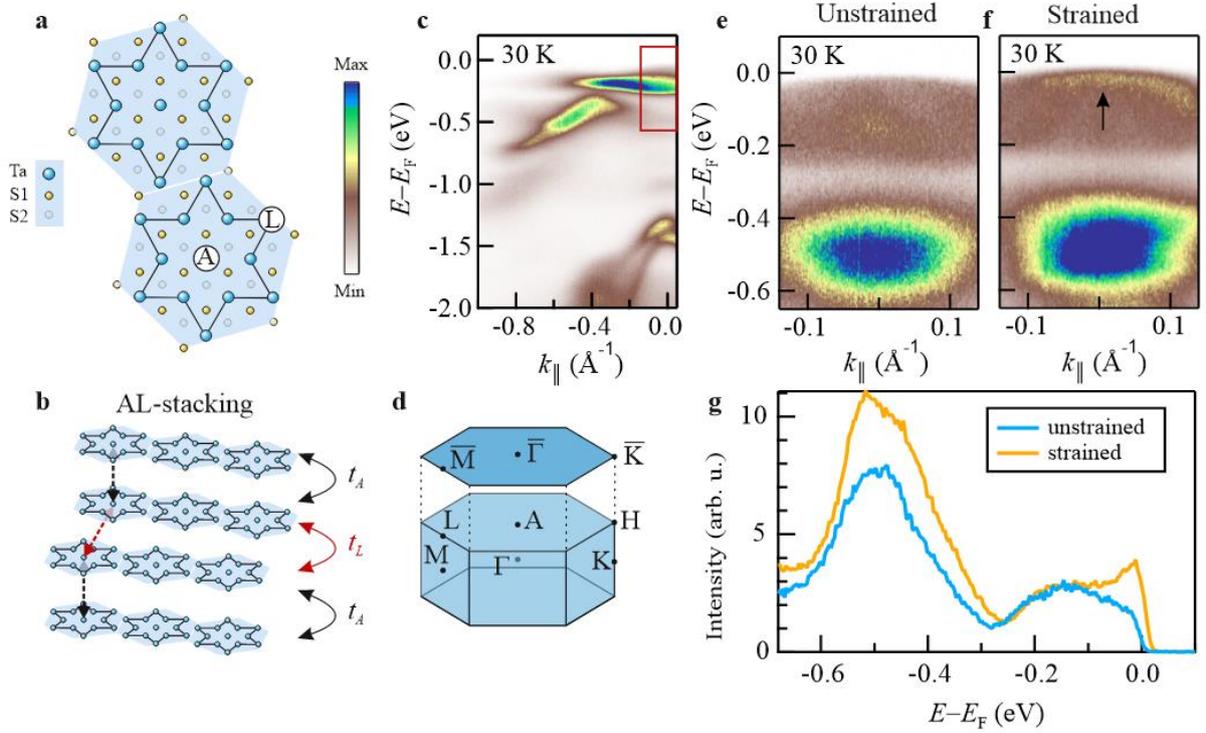

**Fig. 2 Overview of 1$T$-TaS$_2$ and strain-induced flat band. a** Top-down view of the TaS$_2$ surface highlighting the atoms involved in the "star-of-David" structural rearrangement in the CCDW phase. The central atom "A" carries the Mott electron in a $d_{z^2}$ orbital. The Ta atom in the outer shell is designated "L" in line with the previous convention[32]. **b** The equilibrium stacking arrangement in the CCDW phase: bilayers form with the A-atoms vertically aligned. Between bilayers the A and one of the symmetry equivalent L-atoms are randomly aligned. **c** Photoemission spectra obtained along the $\bar{\Gamma} - \bar{M}$ direction of the surface Brillouin zone shown in (**d**) at 30 K and with 21.2 eV photon energy. The red box highlights the region probed in (**e**) and (**f**). **e**, **f** Photoemission spectra obtained with 6.2 eV photons for an unstrained and uniaxially strained sample under otherwise the same conditions. The strain-induced flat band is highlighted by the red arrow. **g** Line cuts through the data in (**e**) and (**f**) integrated over ±0.036 Å$^{-1}$ around the $\bar{\Gamma}$-point.

We have tracked the evolution of this quasi-particle as a function of temperature in order to ascertain its relation to the CDW behaviour in TaS$_2$. For the unstrained sample (Fig. 3a), the behaviour is consistent with previous reports; namely, the metallic NCCDW phase at room temperature abruptly develops a gap at the CCDW transition, thereby removing spectral weight from $E_F$ below 180 K. After straining the same crystal, the evolution below the transition temperature is markedly different. Instead of remaining gapped, the spectral weight at $E_F$ increases with reducing temperature, with the quasi-particle peak becoming evident below 50 K. The raw 2D dispersion data at selected temperatures are presented in the Extended Data Figs. 3 and 4. The contrast between strained and unstrained samples is highlighted in Fig. 3c, where the spectral weight integrated above $E_F$ at each measured temperature is presented. In the strained sample, a small increase of spectral weight at $E_F$ can already be seen at 150 K, but becomes considerably enhanced as the sample is cooled to 15 K. The trend suggests that spectral weight may increase further at lower temperatures. Although the transition temperature itself appears to be unchanged upon application of strain, changes in the spectral weight distribution are already evident directly below the transition at 170 K (see Extended Data Fig. 5). This clearly implies that the quasi-particle behaviour sets in only in the CCDW phase, where Mott physics has previously been postulated. The emergence of a narrow metallic quasi-particle peak is clear evidence for strongly correlated physics: in the range where the kinetic energy term ($t$) and electronic interactions ($U$) in a Mott system become roughly equal, a narrow quasi-particle band appears at $E_F$ and gradually loses coherence with increasing temperature. Conversely, no additional peak is expected in the gap of a band insulator upon external perturbation. The $U/t$ ratio in a correlated system can be altered either by modifying the Coulomb interaction strength or the bandwidth, the latter being potentially controllable via uniaxial strain[18,20]. However, the question of what causes the transition into the correlated metal phase remains open.

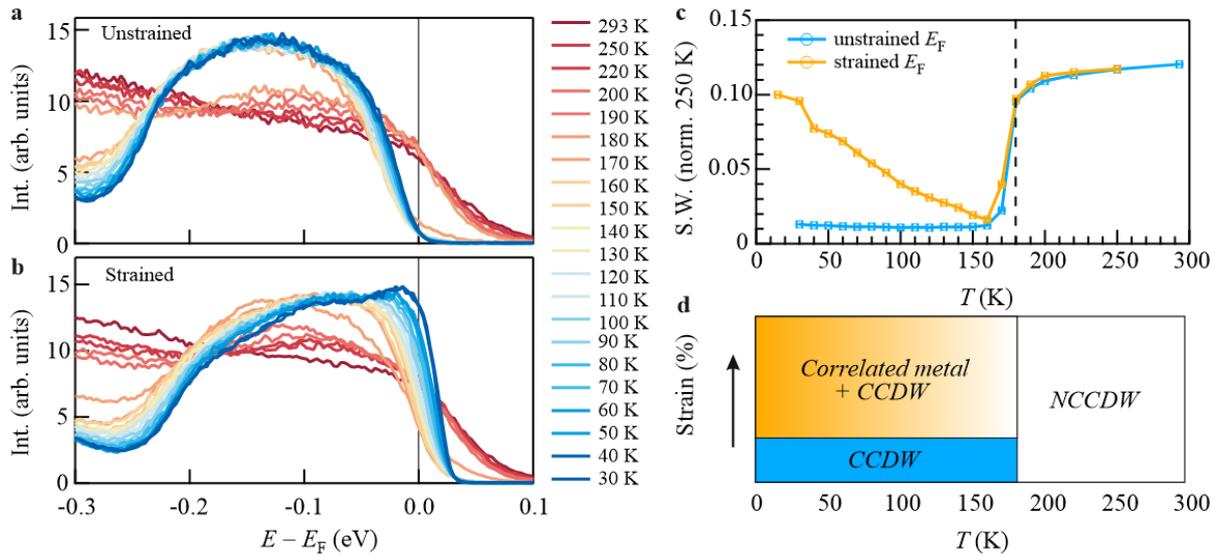

**Fig. 3 Evolution of the quasi-particle peak with temperature. a, b** Line cuts through the photoemission spectra integrated around the $\bar{\Gamma}$-point across the measured temperature range for (**a**) an unstrained and (**b**) a strained sample. Note that the data are acquired on a different sample compared to that measured in Fig. 2. **c** Spectral weight as a function of temperature. The curves are obtained for intensity integrated over [0 : 0.1] eV. The dashed line is the NCCDW to CCDW transition temperature. **d** Schematic phase diagram for strain and temperature.

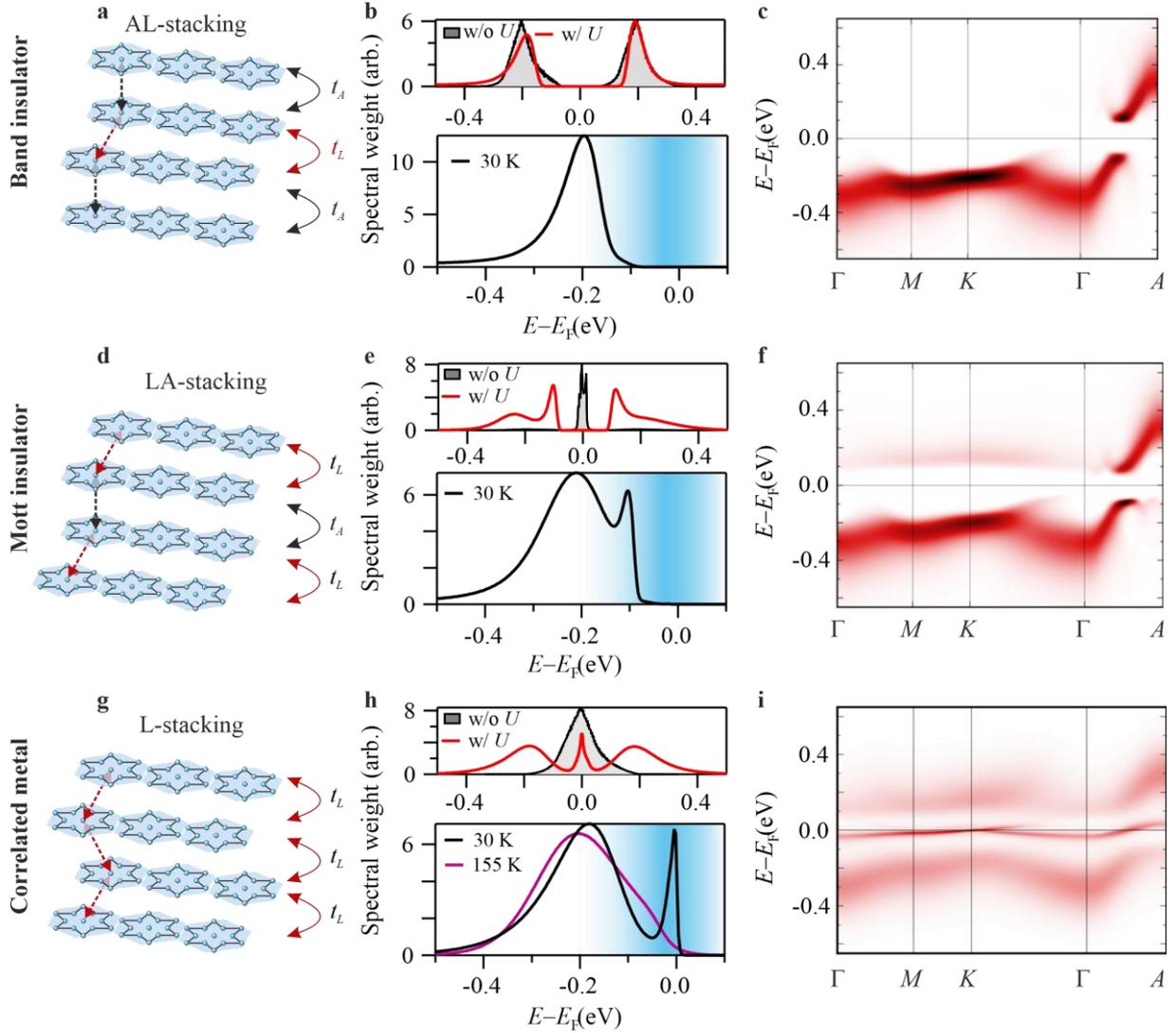

**Fig 4 Electronic behaviour for different stacking arrangements. a** The ground-state stacking previously discussed. **b (upper)** Electronic density of states in the top TaS$_2$ layer integrated over the in-plane Brillouin zone without (grey) and with (red) electronic correlations included, highlighting the band insulating behaviour for this arrangement; **(lower)** integrated density of states from the first three TaS$_2$ layers multiplied by the Fermi distribution at 30 K. **c** Calculated spectral function $A(k,\omega)$ over the full Brillouin obtained via the Fourier transform of the relevant real-space dispersions. **d, e, f** The same calculations as in (**a**)-(**c**) for the LA-stacked surface. In this case, the non-interacting structure is metallic (**e, upper**) and a Mott gap is opened by electronic interactions (**e, lower**). The Mott gap is smaller than the hybridisation gap for the band insulating AL-stacking. **g, h, i** The same calculations as in (**a**)-(**c**) for the fully L-stacked system, previously shown to be close in energy to the ground state AL-stacking. The metallic quasi-particle appears at 30 K (**h, lower**), and is washed out at 155 K, in line with the measured data. The spectral function (**i**) reveals the flat band throughout the Brillouin zone. Notably the flat band appears at $E_F$ at 1/3$^{rd}$ of the out-of-plane $\Gamma-A$ distance, compatible with the $k_z$ position expected for 6.2 eV photon energy [49].

To gain further insight into the mechanism involved in the strain-driven transition, we performed *GW* plus extended dynamical mean-field theory (*GW* + EDMFT) calculations on a semi-infinite stack of TaS$_2$ layers (see Methods) as summarised in Fig. 4. The electronic structure in the lowest energy configuration (Fig. 4a), termed "AL-stacking" [32] is characterised by two inequivalent interlayer hopping values between the $d_{z^2}$ orbitals: $t_A = 0.2$ eV within the bilayer where the stacking vector is **T**s = c; and $t_L = 0.045$ eV between neighbouring bilayers with **T**s= -2**a**+2**c**. The structure is known to be insulating with a hybridisation gap caused by hopping within the bilayer [31,32], and this property is reproduced by our density of states and momentum-resolved spectral function reported in Figs. 4b and c respectively. A second possibility in the AL configuration is to terminate the surface within a bilayer [33,34] (Fig. 4d). As has been recently shown this "LA" stacked termination is also insulating, however in this setup it is strong correlations that open a Mott gap in an otherwise metallic surface state [44] (Fig. 4e). The gap size in the local spectrum of the near-surface region is smaller in this case but the structure remains insulating along all momentum directions (Fig. 4f). Clearly neither of these equilibrium structures (AL- or LA-stacking) can explain the emergent metallic band in strained samples, suggesting the need for a structure where the interlayer hopping is modified.

Since strain adds mechanical energy to the system, we consider the second lowest energy structure [32], with an interlayer stacking as shown in Fig. 4g. To study this structure we modify the interlayer hopping such that there is equivalent coupling between all layers (see Methods). This "L-stacking" structure is very close in energy to the AL stacking structure, and has been suggested to become the ground state under pressure [32]. Remarkably, the *GW*+EDMFT calculations (Fig. 4h) reveal the appearance of a narrow quasi-particle peak at $E_F$ plus an incoherent background, which qualitatively resembles the structure observed in our ARPES measurements in Figs. 2 and 3. In contrast, the density of states predicted by a non-interacting calculation (Fig. 4h) is not consistent with the experimental findings, which critically highlights the importance of correlations in this emergent metallic state. This is an important result as, the role of correlations in the complex phase diagram of TaS$_2$ has long been debated. Further evidence for the role of correlations is obtained by increasing the temperature in the *GW*+EDMFT calculation to 155 K. The fact that this removes the quasi-particle peak further highlights the origin of this state in strongly correlated physics, as a non-interacting metallic band cannot be so easily removed by heating. As is evident from Fig. 4i, the metallic band disperses very narrowly with a bandwidth of only 50 meV at 30 K, comparable to that observed in experiment. The quasi-particle extends into the bulk, as seen in the layer-resolved calculation shown in Extended Data Fig. 6. We note that this bulk correlated metal in the L-stacking structure contrasts with the correlated Mott gap in the LA structure which only appears at surfaces or stacking defects.

The fact that the calculated quasi-particle is more clearly separated from the incoherent background than in the experimental data suggests that the strain-induced transition is spatially inhomogeneous and depends on the local strain distribution. This would result in a superposition of signal from both AL and L-stacked structures within the region probed by our laser spot. This is further supported by the data in Fig. 2 and 3 which, although qualitatively similar, show differing relative heights of the quasi-particle peak compared to the background. These independent data sets were obtained on separate crystals and strain devices which suggests that the local strain distribution were different. Nano-ARPES may help to better isolate the phases and their spatial distributions.

**Discussion**

Flat bands are typically associated with strong correlation physics due to the high density of states and small kinetic energy intrinsic to such states (*t*<<*U*), as recently demonstrated in Kagome crystals [50–52] and twisted heterostructures [5,6]. A flat metallic $d_{z^2}$ band in TaS$_2$ has been predicted previously by DFT calculations as a result of spin-orbit coupling, and was postulated as the origin of the Mott phase [42]. However, as discussed above, in the AL stacked ground state the system is already gapped due to the large interlayer hopping ($t_A = 0.2$ eV, evident from the bandwidth along $\Gamma$-A in Fig. 4c) which causes hybridisation and band insulating behaviour. While correlations are still present in this structure, they are not responsible for the gap formation and therefore the *U* is ineffective despite a value of 0.4 eV.

In stark contrast, the interlayer hopping is significantly reduced to only $t_L$ = 0.035 eV in the L-structure, thereby removing the hybridisation gap (Fig. 4h). This allows the latent Mott correlations to become dominant within the TaS$_2$ layers. The in-plane bandwidth in the L-stacked structure is around 0.2 eV (Fig. 4h), comparable to $U$=0.4 eV, thereby placing the system in the correlated quasi-particle regime of Fig. 1. It is worth emphasising that the L-structure is indeed a correlated metal, not a Mott insulator, due to the large in-plane bandwidth present in this structure. In contrast, the in-plane bandwidth at the surface of the other potentially metallic structure (with LA stacking) is only 0.05 eV (Fig. 4e), placing it deep into the strongly correlated regime ($t<<U$) and explaining the correlated insulator behaviour [33,34,44]. We remark that the kinetic energy values in our calculations for AL and LA stacking were obtained with reference to previous STM data [33,34]. In the L-stacking structure we further constrain this value via the temperature dependence by requiring that the quasi-particle disappears at 155 K in order to be compatible with our experimental observations. To summarize, the emergence of the narrow band at $E_F$ (Fig. 4i) occurs due to the reduced interlayer coupling in L-stacking compared with the equilibrium AL stacking. Our measurements therefore highlight the role played by interlayer coupling as critical to stabilising correlated behaviour and the ability of uniaxial strain to modify both. We note that a similar argument regarding the increased relevance of Mott correlations in an altered stacking arrangement was suggested as an explanation for an insulator-to-insulator transition observed during heating from the CCDW phase [53].

Given the appearance of the correlated metal only below the CCDW transition temperature, it is highly probable that the partial structural rearrangement into the L-structure occurs during the cooling transition from the NCCDW phase at 180 K in the strained sample. According to previous theoretical reports[32], the energy difference between AL and L-stacking is only 1.1 meV per star-of-David and the two structures even become degenerate under 0.3% reduced inter-layer spacing e.g. under pressure. The energy difference is, however, likely underestimated, as this would imply a significant proportion of L-stacking already at temperatures above 10 K, which has not been reported. This logic suggests an energy difference between structures closer to 20 meV, which therefore sets the scale for strain induced changes in the present experiment. Regardless of the exact energy values, the fact that the AL and L structures are much closer in energy than all other possible stacking orders, combined with the reversal of the energy order under external pressure, supports this as the physical mechanism responsible for the formation of the flat band. This motivates future theoretical investigations of strained TaS$_2$ samples to better understand the detailed energy landscape induced by strain, as well as X-ray diffraction to probe the *c*-axis structure. Similar considerations are very likely relevant for the Kagome series $AV_3Sb_5$, where the stacking of star-of-David structural distortions plays may also play a key role in the energetics of charge ordering[54,55]. The agreement between our data and calculations strongly implies that, by favouring the L-stacking formation during the NCCDW to CCDW phase transition, strain modifies the *c*-axis structure of the CCDW phase. It is therefore possible that, at the strain levels applied here, different behaviour would be observed for crystals cooled below $T_{CCDW}$ and then strained. It is worth noting that the energy gain through formation of the CCDW phase is on the order of 200 meV per Ta atom[42,46], considerably larger than the estimated changes caused by strain at the current level, meaning we do not expect changes to the in-plane CCDW structure. Given the L-stacking is a bulk order, and since strain transmits through the entire crystal in our geometry, we would expect changes to be visible in resistivity measurements, and suggest to look for the emergence of superconductivity or other correlated phases in strained samples.

Intriguingly the behaviour observed under strain does not appear identical to that under hydrostatic pressure[23,56]. Measurements in the CCDW phase under increasing pressure show that the structure initially transitions to the NCCDW phase followed by the onset of superconductivity at higher pressures. Our present results in low-temperature strained samples do not resemble the band structure in the NCCDW phase, despite the fact that both are metallic (see Extended Data Figs. 3 and 4). Nevertheless, the increased relevance of correlations in the strained L-structure along with the high density of states associated with the flat band may still be relevant to understanding the appearance of superconductivity in TaS$_2$. Additionally, the strain-induced scenario outlined above provides a potential mechanism for the appearance of the metastable hidden phase. STM studies have shown that changes of the interlayer registry can indeed produce local metallic behaviour[25,33,57], while X-ray diffraction has revealed that the dimerised layers of the CCDW phase can be removed by the application of a single picosecond optical pulse[58]. We hypothesise that strain introduced by an excitation may be responsible for the emergence of the metastable phase. In general, pulsed optical excitation can produce transient electronic temperatures of the order $10^3$ kelvin, which can in turn induce significant atomic movement due to the altered population of bonding and anti-bonding states[59]. In the case of TaS$_2$, optical or electronic pulses may produce a similar non-equilibrium electronic landscape which would result in shifted atomic positions i.e. an effectively strained lattice. Thereafter, dynamic cooling in this strained energy landscape could lead to the stacking configuration observed in experiments. This motivates future work on the dynamic nature of the interlayer behaviour, both experimentally and theoretically. Independent of this discussion, our results expose the possibility of using uniaxial strain to create alternative protocols for inducing hidden phases.

In summary, we have experimentally demonstrated that uniaxial strain induces a metallic flat band in 1*T*-TaS$_2$, which appears in the CCDW phase. The appearance of such a quasi-particle state within the gap is a clear indication for strong electronic correlations and Mott physics in strained TaS$_2$, with direct relevance for the microscopic mechanisms governing superconducting and metastable hidden phases in this archetypal material. The supporting *GW*+EDMFT calculations confirmed that a modification of interlayer coupling, linked to the stacking arrangement in the strained crystal, and an associated bandwidth-driven insulator-metal transition results in a strongly-correlated metal state with a renormalized quasi-particle band and incoherent Hubbard band features. Our observations validate strain as a parameter to control correlated behaviour via interlayer coupling and highlight the potential of this approach for manipulating phase diagrams in layered systems.

**Materials and Methods**

*Crystal growth* – High quality single crystals of 1*T*-TaS$_2$ were grown by the chemical vapour transport (CVT) method using TaCl$_5$ as a transport agent. First, polycrystalline samples of 1*T*-TaS$_2$ were synthesized by mixing stoichiometric amounts of tantalum (powder, Alfa Aesar, 99.99 %) and sulphur (pieces, Alfa Aesar, 99.999 %) of a total mass of 1 g. The reactants were sealed in a quartz ampoule (l = 9 cm, Øl = 9 mm) under 1/3 atm. The quartz ampoule was heated to 1000 °C at a rate of 180 °C/h, kept at this temperature for 4 days and subsequently quenched in water. 200 mg of the as synthesized powder and 10 mg TaCl$_5$ (anhydrous, powder, VWR, 99.999%) were sealed in a quartz ampoule (l = 20 cm, Øl = 7 mm) under vacuum and heated for 6 days in a two zone furnace in which the source zone and the growth zone temperatures and were fixed at 980 °C and 850 °C, respectively. Eventually, the tube was quenched in cold water to ensure the retaining of the 1*T* phase. The product was confirmed to be phase pure by powder X-ray diffraction (PXRD). Patterns were collected on an STOE STADI P diffractometer in transmission mode equipped with a Ge-monochromator using Cu K$_{\alpha 1}$ radiation and on a Rigaku SmartLab in reflection mode using Cu K$_\alpha$ radiation. Transport measurements from 300 to 10 K were carried out with an excitation current of I = 1 mA on a Quantum Design Physical Property Measurement System (PPMS) Evercool with a 9 T magnet. Gold wires (25 µm) were connected to the sample with silver epoxy in the standard four probe method.

*Sample strain* – Uniaxial strain was applied using a modified version of a previously described home built strain cell[20], from which it was possible to reach 0.2% strain as measured by an electronic strain gauge in a Wheatstone bridge configuration. Samples to be prepared for straining were chosen to have large flat areas with minimal cracks or flakes at the surface as viewed under an optical microscope, in order to allow for more homogeneous strain application. Orientation of crystals was carried out using a commercial Laue diffractometer. Crystals were strained parallel to the *a** axis of the room temperature phase. Bulk samples were initially cleaved with a scalpel to remove thicker layers, and then mounted onto the unstrained device and further thinned by Scotch tape cleaving. Unstrained and strained comparisons presented in this work were obtained on the same crystal before and after straining. The data presented in Figs. 2 and 3 were obtained on different crystals.

*ARPES* – Photoemission measurements were carried out in a base pressure of $10^{-11}$ mbar. VUV photons at 21.2 eV were produced using a monochromatised He plasma source (SPECS GmbH). Low-energy UV photons were generated using a commercial optical setup (Harmonix, APE GmbH) generating tuneable output in the range 5.7–6.3 eV in nonlinear crystals. Harmonic generation was driven by the output of a tuneable OPO pumped by a 532 nm Paladin laser (Coherent, Inc.) at 80 MHz. The beam polarization used was linear horizontal (*p*-pol) and the beam size was 80 × 80 μm$^2$. The sample surface was scanned by the encoded motion of a 6-axis cryogenic manipulator (SPECS GmbH). Spectra were acquired using a Scienta-Omicron DA30 analyser.

*Electronic structure calculations* – The *GW*+EDMFT simulations were performed with the set-up described in Ref. [44], which considers eight 1*T*-TaS$_2$ layers and an embedding potential for the bulk. The local correlation and screening effects are treated with extended dynamical mean field theory (EDMFT) and the nonlocal ones at the *GW* level. We use a realistic band structure for the in-plane hopping[60] and interlayer hoppings $t_A = 0.2$ eV and $t_L = 0.045$ eV which were optimized by comparing the local spectrum of the surface layer in the AL and LA stacked systems to the corresponding STM spectra [33]. In the L-structure we use $t_L = 0.035$. Upon convergence, the GW+EDMFT calculation provides a momentum-and frequency-dependent self-energy $\sum_{ab}(k_{||}, i\omega_n)$ and lattice Green's function $G_{ab}(k_{||}, i\omega_n)$, with *a*, *b* layer indices and $k_{||}$ the in-plane momentum. The local spectral function $A_a = -\frac{1}{\pi} Im \frac{1}{N_{k_{||}}} \sum_k G_{aa}(k_{||}, \omega)$ is obtained with maximum entropy analytical continuation. The momentum-resolved spectra were obtained by Fourier-transforming the layer-resolved spectra with respect to the layer index [44].

**Acknowledgements**

C.W.N., B.S. M.R, G.K. and C.M. acknowledge support from the Swiss National Science Foundation Grant No. P00P2_170597. F.P. and P.W. acknowledge support from the Swiss National Science Foundation through NCCR MARVEL and from the European Research Council through ERC Consolidator Grant 724103. C.W. and F.O.vR acknowledge the support from the Swiss National Science Foundation under grant no. PCEFP2_194183 and by the Swedish Research Council (VR) through a neutron project grant Dnr. 2016-06955. The calculations were performed on the Beo05 clusters at the University of Fribourg and the Piz Daint cluster at the Swiss National Supercomputing Centre (CSCS) under project ID mr26. We gratefully acknowledge M. Wolf and L. Rettig for providing constructive feedback, J. Chang for access to the Laue diffractometer at the University of Zurich, and Q. Wang for technical support. Skilful technical support was provided by M. Andrey, B. Hediger, G. Bächler and F. Bourqui.

**Contributions**

C.W.N. conceived the project. C.W. and F.O.v.R fabricated the samples and performed the resistivity and diffraction measurements. ARPES and strain measurements were performed by C.W.N. and B.S. with assistance from M.R. and G.K., and were analysed by C.W.N. Electronic structure calculations were carried out by F.P. and P.W. The project was managed by C.W.N. together with C.M. C.W.N. wrote the manuscript with input from all authors.

*nicholson@fhi-berlin.mpg.de

# Extended Data for:

# Emergent flat band under uniaxial strain reveals Mott correlations in 1$T$-TaS$_2$

Christopher W. Nicholson[1,2,*], Francesco Petocchi[1], Björn Salzmann[1], Catherine Witteveen[3,4], Maxime Rumo[1], Geoffroy Kremer[1], Fabian O. von Rohr[3,4], Philipp Werner[1], Claude Monney[1]

[1]*University of Fribourg and Fribourg Centre for Nanomaterials, Chemin du Musée 3, CH-1700, Fribourg, Switzerland*

[2]*Fritz Haber Institute of the Max Planck Society, Faradayweg 4-6, D-14195, Berlin, Germany*

[3]*Department of Physics, University of Zurich, Winterthurerstrasse 190, CH-8057 Zurich, Switzerland*

[4]*Department of Quantum Matter Physics, University of Geneva, 24 Quai Ernest-Ansermet, CH-1211 Geneva, Switzerland*

**Extended data Fig. 1 Crystal characterisation: transport & diffraction**

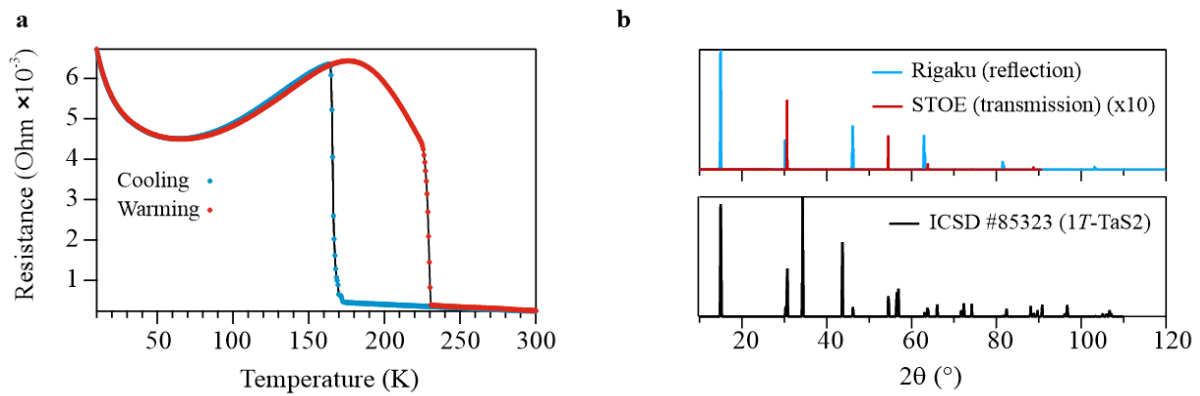

**a** Resistance curve as a function of temperature during a cooling/warming cycle. The NCCDW to CCDW transition is clearly evident, as is the typically observed hysteresis upon warming. The resistance was measured from 300 to 10 K with an excitation current of I = 1 mA on a Quantum Design Physical Property Measurement System (PPMS) Evercool with a 9 T magnet. Gold wires (25 µm) were connected to the sample with silver epoxy in the standard four probe method. **b** (Upper curves) X-ray diffraction peaks obtained on a single crystal in both reflection and transmission geometries. Patterns were collected on an STOE STADI P diffractometer in transmission mode equipped with a Ge-monochromator using Cu K$_{\alpha 1}$ radiation and on a Rigaku SmartLab in reflection mode using Cu K$_\alpha$ radiation. (Lower curves) Reference spectra #85323, corresponding to the 1$T$-phase of TaS$_2$, from the Inorganic Crystal Structure Database (ICSD).

**Extended data Fig. 2 Bandwidth at $E_F$**

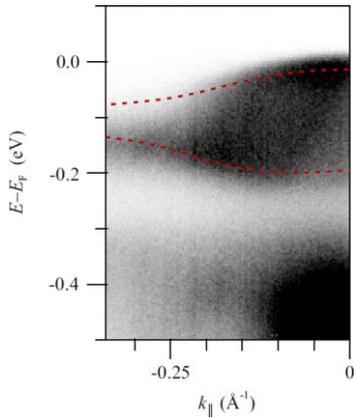

ARPES data of the strained crystal rotated away from the G-point towards M. From the dispersion we estimate a bandwidth of 70 meV for the band close to $E_F$. Red dashed lines are guides to the eye.

**Extended data Fig. 3 ARPES temperature comparison obtained with He lamp**

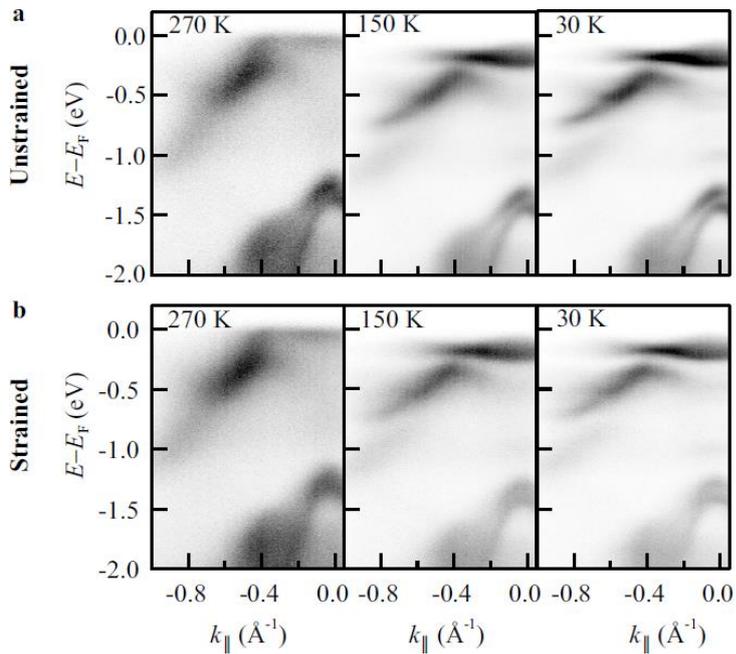

**a** ARPES spectra at selected temperatures obtained in the $\Gamma$-M plane at 21.2 eV in an unstrained crystal at 270 K (NCCDW phase), 150 K (CCDW phase) and 30 K. **b** The corresponding ARPES spectra obtained in the same crystal under strain.

**Extended data Fig. 4 ARPES temperature comparison obtained with laser (6.2 eV)**

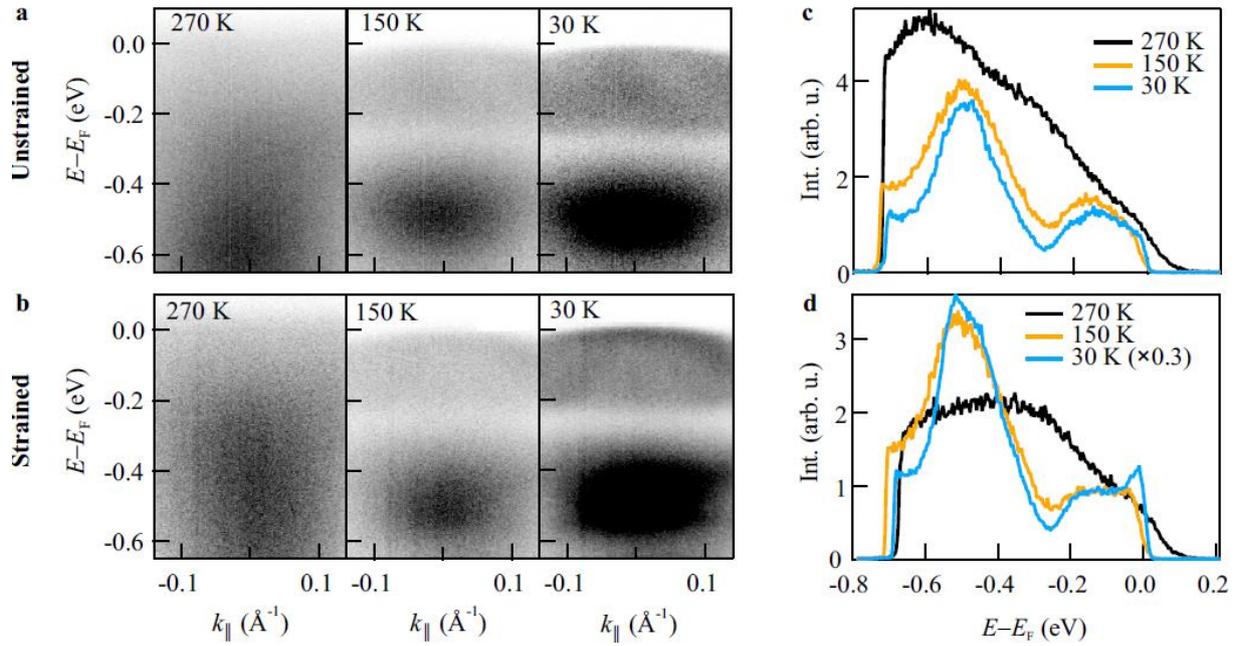

**a** ARPES spectra at selected temperatures obtained in the Γ-M plane at 6.2 eV in an unstrained crystal at 270 K (NCCDW phase), 150 K (CCDW phase) and 30 K. **b** The corresponding ARPES spectra obtained in the same crystal under strain. The data are obtained on the same crystal as Extended Data Fig 3. **c** Line cuts through the unstrained data and **d** through the strained data, highlighting the emergence of the narrow quasi-particle band at $E_F$.

**Extended data Fig. 5 Temperature curve integrated over the quasi-particle**

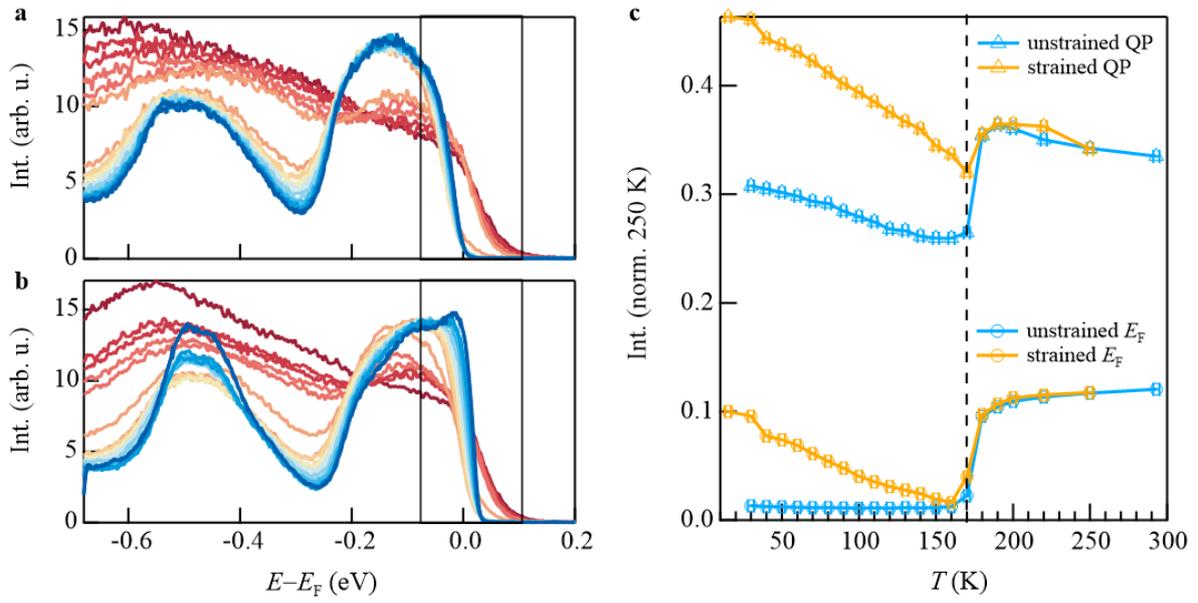

**a** Line cuts through the photoemission spectra integrated around the $\bar{\Gamma}$-point across the measured temperature range for an unstrained crystal. Note the extended energy range compared with Fig. 3 of the main text. **b** The same temperature dependent line cuts for the same crystal under strain. **c** Spectral weight as a function of temperature. The upper curves are obtained for intensity integrated over [-0.75 : 0.1] eV i.e. the full quasi-particle (QP). A difference in behaviour between strained and unstrained crystals is already evident at 170 K. The lower curves are the same as shown in the main text Fig. 3. The dashed line is the NCCDW to CCDW transition temperature.

**Extended data Fig. 6 Layer resolved calculation**

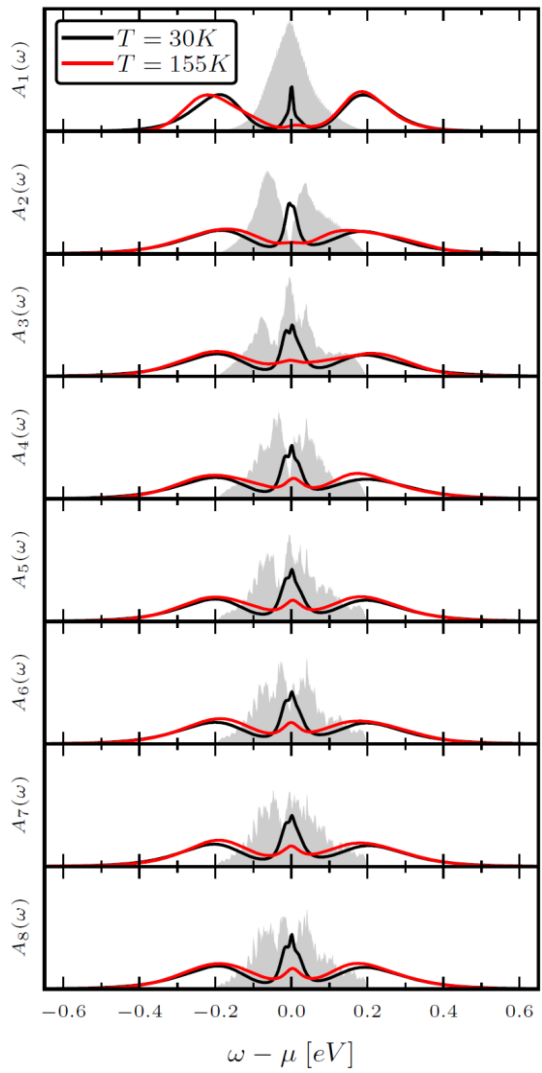

Calculated spectral function $A_n(\omega)$ as a function of the layer index $n$ for the L-type stacking described in the main text, with $n$=1 being the surface layer. Grey solid curves are the non-interacting spectra obtained with DFT using the LDA. Black and red curves are the interacting spectra calculated with *GW*+EDMFT and are obtained at 30 K and 155 K respectively. The metallic quasi-particle peak is observed to extend into the bulk.